\documentclass[english,twocolumn,twocolumn,showpacs,preprintnumbers,amsmath,amssymb,superscriptaddress]{revtex4}
\usepackage[T1]{fontenc}
\usepackage[latin1]{inputenc}
\usepackage{graphicx}

\pdfoutput=1
\makeatletter

\makeatletter

\makeatletter

\usepackage{color}

\usepackage{dcolumn}

\usepackage{bm}

\def\lsim{\mathrel{\rlap{\lower4pt\hbox{\hskip1pt$\sim$}}
    \raise1pt\hbox{$<$}}}         
\def\gsim{\mathrel{\rlap{\lower4pt\hbox{\hskip1pt$\sim$}}
    \raise1pt\hbox{$>$}}}         

\usepackage{babel}
\makeatother
\begin{document}

\title{
Anisotropic emission of direct photons from Au+Au collisions at $\sqrt {s_{\rm NN}}=200~{\rm GeV}$ with EPOS3 and IP-Glasma }

\author{Fu-Ming Liu}

\affiliation{Institute of Particle Physics, Central China Normal University, Wuhan,
China}

\author{Sheng-Xu Liu}

\affiliation{Institute of Particle Physics, Central China Normal University, Wuhan, China}

\affiliation{ Wuhan Foreign Languages School, Wuhan, China}

\author{Klaus Werner}

\affiliation{Laboratoire SUBATECH, University of Nantes - IN2P3/CNRS - Ecole desMines,
Nantes, France}

\date{\today}

\begin{abstract}
The observed large elliptic flow ($v_2$) of direct photons in relativistic heavy ion collisions challenges us to explain. 
In this paper we consider only two sources of direct photons, prompt photons and thermal photons. 
Prompt photons are calculated with QCD to next leading order, which fits well the spectra at high transverse momentum ($p_{\rm t}$) region. 
Thermal photons are calculated with AMY rate in a 3+1D viscous hydrodynamical evolution of the collision systems. 
Two types of initial conditions, EPOS3 and IP-Glasma, are employed.  
Our results with IP-Glasma initial condition agree with results in published work in most cases, 
and the difference tells the regions effected by the absence of viscous correction in photon emission rate. 
Because both models are well constrained with hadron data, the hydrodynamic evolution of collision systems in the two models have many common.  
However, the transverse velocity of the collective motion are quite different. 
This makes a big difference in the anisotropic emission of direct photon emission. 
The observed large elliptic flow of direct photons is reproduced with EPOS3 hydrodynamic evolution.   
\end{abstract}
\maketitle

\section{Introduction}
Direct real photons serves as the most versatile tools to study relativistic heavy ion collisions because they are produced by various mechanisms during the whole space-time history of the interacting system. The observation of an unexpectedly large direct photon flow $v_2$ (essentially the same magnitude as hadrons) is a big challenge and so far eludes full and coherent explanation\cite{PHENIX:2011oxq, PHENIX:2015igl}.
It is believed that thermal photons are emitted during the whole evolution of the collision systems, especially stronger at higher temperature at early stage, ie, in QGP phase, but large $v_2$ usually indicates late production at hadronic phase.  
That's why it is a big challenge to simultaneously explain the experimental data of hadrons and photons, such as $p_{\rm t}$ spectra, $v_2$ and $v_3$ and so on. 

Predictably, many theorists took up this challenge, i.e., conferences and dedicated workshops\cite{David161, David309, David310, David311, David312, David313, David314}. 
A review have been given in\cite{David:2019wpt}, where the $v_2$ of direct photons predicted by  hydrodynamic models\cite{Gale:2014dfa}
 and transport models\cite{vanHees:2011vb,Linnyk:2013wma} is lower than experimental data.
Additional sources of direct photons are investigated at different stages with different processes. First, from Glasma, the pre-equilibrium stage \cite{Chiu:2012ij, Monnai:2014kqa}.
 Since quarks becomes substantial only at later stage of Glasma, a delayed chemical equilibrium is helpful to increase photon $v_2$~\cite{Monnai:2014kqa}, but not big enough to explain data\cite{Gale:2021emg}. 
A special photon source from Glasma due to gluon fusion with a triangular Feymann diagram is investigated \cite{Jia:2022awu}. 
This source gives a visible $v_2$ of produced photons, but not big enough to explain the data.
And, this gluon fusion process should be confirmed by the experiments of hadronic collisions where the background is much cleaner than heavy ion collisions. 
By the way, gluon fusion process is also considered in the presence of a magnetic field\cite{Ayala:2022zhu}.
In fact, the correction to photon emission due to the existence of external electro-magnetic (EM) fields excited by the spectator nuclei of the incoming beam, has been investigated, 
where the mean magnetic field strength has been extracted from direct photon data\cite{Sun:2023rhh}.
Large $v_2$ of direct photons may imply an additional source at very late stage, for example,
a special process where a quark and an antiquark binds into a meson state accompanying a photon emission is introduced at hadronization stage~\cite{Fujii:2022hxa}. 
This photon source can increase quite a lot photon $v_2$, if this process can be confirmed experimentally. 

For the purpose of simultaneous explanation to the experimental data of hadrons and photons, successful event generators which can explain already hadronic 
data are employed to calculate photon production. 
We first take EPOS3, where a 3+1D viscous hydrodynamics has been employed to describe the collective motion of the colliding system\cite{Werner:2013tya, Werner:2010aa}. This space-time evolution of colliding system provides a convenient condition to calculate thermal photon emission. 

In order to understand the photon emission from EPOS3, we do exactly the same calculation except replacing initial condition with IP-Glasma\cite{Schenke:2012hg, Schenke:2012wb}.
The IP-Glasma initial condition in a 3+1D viscous hydrodynamic evolution has been well investigated \cite{Gale:2021emg, Paquet:2015lta}. Hadronic data have been well explained, however $v_2$ of direct photons is lower than experimental data. In fact the effects of chemical equilibrium time \cite{Gale:2021emg}, switch initial flow time\cite{Gale:2012rq}   and viscosity correction\cite{Paquet:2015lta}, to photon emission have also been investigated, very meticulous and comprehensive. 
Model comparison will help us understand the results.

\section{Approach}
Prompt photons and thermal photons are considered as the sources of direct photons in our calculation.

The transverse momentum spectra of prompt photons in nucleus-nucleus
collisions are calculated as \cite{Owens1987} 
\begin{eqnarray}
\frac{dN^{\rm pr}} {dy\, d^{2}p_{\rm t}} =  
T_{AB}(b) \{ \sum_{{\displaystyle ab}}
\int dx_{a}dx_{b}G_{a}(x_{a},M^{2})   \nonumber \\
G_{b}(x_{b},M^{2}) \frac{\hat{s}}{\pi}   
 \delta(\hat{s}+\hat{t}+\hat{u})[\frac{d\sigma}{d\hat{t}}(ab\rightarrow\gamma+X) \nonumber  \\
  +  K\sum_{c}\frac{d\sigma}{d\hat{t}}(ab\rightarrow cd)\int dz_{c}\frac{1}{z_{c}^{2}}D_{\gamma/c}(z_{c},Q^{2})]  \}
\label{eq:prompt}
\end{eqnarray}
 where 
$T_{AB}(b)$ is the nuclear overlapping function at an impact
parameter $b$ for each centrality, same as our previous work~\cite{Liu:2008eh},
$G_{a}(x_{a},M^{2})$ is parton distribution functions (PDF)
in proton, the elementary processes $ab\rightarrow\gamma+X$ are Compton
scattering $qg\rightarrow\gamma q$ and annihilation $q\bar{q}\rightarrow g\gamma$
and the second term covers high order contribution with photon fragmentation
functions $D_{\gamma/c}(z_{c},Q^{2})$ being the probability for obtaining
a photon from a parton $c$ which carries a fraction $z$ of the parton's
momentum. In our calculation, MRST2001 \cite{Martin:2001es} PDF
is employed and $K$=2 is used to take into account high order contribution
of hard parton production.

The transverse momentum spectra of thermal photons can be written as 
\begin{equation}
\frac{dN^{\rm th}} {dy\, d^{2}p_{\rm t}} 
=\int d^{4}x\,\Gamma(E^{*},T)\label{eq1}\end{equation}
 with $\Gamma(E^{*},T)$ being the Lorentz invariant thermal photons
emission rate which covers the contributions from the QGP phase \cite{AMY}
and HG phase \cite{MYM}, $d^{4}x=\tau\, d\tau\, dx\, dy\, d\eta_{s}$
being the volume-element, and $E^{*}=p^{\mu}u_{\mu}$ the photon energy
in the local rest frame. Here $p^{\mu}$ is the photon's four momentum
in the laboratory frame, $T$ and $u^{\mu}$ are the temperature and
the local fluid velocity, respectively, to be taken from the hydrodynamic
	evolution of the collision system.

We consider thermal photon radiation from hydro initial time $\tau_0$ till the end whenever the region with energy density 
above $e^{\mathrm{th}}$ in both EPOS and IP-Glasma initial conditions. 
Local thermal equilibrium and chemical equilibrium are assumed since $\tau_0$ in our calculation. 
Due to the existence of dissipative forces in the viscous hydro evolution, the phase space distribution of partons/hadrons has a deviation from the distribution in full (thermal, chemical and mechanical) equilibrium. 
The deviation can be estimated with the Boltzmann equation in a relaxation time approximation. Thus photon emission rate should be effected in both QGP and HG phase. 
At present, not all processes of photon production in QGP and HG phase are amenable to a calculation of viscous (shear and bulk) corrections.
For example, the AMY rate covers the processes of all orders according to the hard thermal loop calculation~\cite{AMY}, 
while it is convenient to do viscous correction to $2 \rightarrow 2$ processes, as list in~\cite{Paquet:2015lta}. 
Taking into account the uncertainty of the chemical equilibrium time and the temperature-dependence of viscosity, viscous correction to photon emission rate is ignored in our calculation.

In our calculation of thermal photon emission, the space-time evolution of collision systems 
is governed by the 3+1D viscous hydrodynamic equations.
The numerical solution\cite{Werner:2013tya} in EPOS3102 with equation of states is kept the same for both initial conditions.
In order to calculate thermal photon production, we keep the hydrodynamic evolution till $e^{\mathrm{th}}=0.08~{\rm GeV}/fm^3$.

EPOS initial condition is based on event generators\cite{NEXUS,VENUS} 
in a relatively long history of constraining model parameters with experimental data 
from electron-positron annihilation, lepton-nucleon scattering, hadron-hadron interaction, hadron-nucleus interaction and nucleus-nucleus interaction at various energies.   
For Au-Au collisions at $\sqrt{s_{\rm NN}}$=200~GeV, multi-pomeron exchange occurs, where hard pomerons obey perturbative QCD and soft pomerons follow the phenomenological Gribov-style treatment used in the pre-QCD time.
Then pomeron information is turned into color tubes with size and energy density distributed in the space. This becomes the hydro initial condition of EPOS. The initial transverse flow at $\tau_0=0.35~fm/c$ is assumed to be zero. 
    
In the case of IP-Glasma\cite{Schenke:2012wb,Schenke:2012hg}, the fluctuating color charges of the colliding nucleus are treated as the static sources of the transverse color glass condensate (CGC) fields. 
The CGC fields are determined by solving the classic Yang-Mills equations.
Constructing the energy-momentum tensor of the CGC fields, one can get an initial condition for hydro evolution.  
The mean transverse distribution of the fluctuating color charges involve the saturation scale $Q_s$.  Flow velocity is initialized at $\tau_{\rm switch}=1/Q_s = 0.2~fm/c$, earlier than hydro initial time $\tau_0=0.4~fm/c$. Therefore, when we begin to count thermal photons from $\tau_0$, the initial flow velocity is non-zero.

The triple differential spectrum can be written as a Fourier series,
\begin{eqnarray}
\frac{d^{3}N}{dy\, d^{2}p_{\rm t}}=\frac{d^{2}N}{2\pi p_{\rm t}\, dp_{\rm t}\, dy}\left(1+\sum_{n=1}^{\infty}2v_{n}\cos(n\phi)\right),\label{eqdn}\end{eqnarray}
 where $\phi$ is the azimuthal angle of photon's momentum with respect
to the reaction plane.
In lab, experimentalists have to work hard to reconstruct the reaction plane according to particles detected. 
While in model calculation, it is clearly defined to be the plane containing the impact parameter and beam axis. 
The $n$th flow of thermal photons is quantified by the harmonic coefficient $v_{n}$ 
\begin{eqnarray}
v_{n}(p_{\rm t},y)=\frac{\int d\phi\cos(n\phi)d^{3}N/dy\, d^{2}p_{t}}{\int d\phi d^{3}N/dy\, d^{2}p_{t}}.\end{eqnarray}

It is believed that the flow of prompt photons, $v_n^{\rm pr}=0$. Therefore
the flow of direct photons is calculated as \begin{equation}
 v_n  =  \frac {N^{\rm th}}  { N^{\rm th} +N^{\rm pr} } * v_n^{\rm th}  
\label{normalize}
 \end{equation}   
where 
 $ N^{\rm th} $ and  $N^{\rm pr} $ stands for the number of thermal photons and prompt  in each given $p_t$ bin, respectively.  

Here we discuss the origin of $v_n$ which describe the anisotropic emission of photons. 
According to Eq.~(\ref{eq1}), the momenta of emitted photons are isotropically distributed in the local rest frame.
We observe photons in the laboratory frame, while the local rest frame moves at the speed $u_\mu$ with respect to the  laboratory frame. 
Here $u_\mu$ is fluid velocity obtained from solving hydrodynamic equations of the collision system.
To get the momenta of emitted photons in the laboratory frame, Lorentz transformations with fluid velocity should be done.
So the eccentricity of fluid velocity, makes the anisotropic emission of thermal photons directly.
The space eccentricity causes azimuthal anisotropies in transverse pressure gradients which push the fluid to have the collective motion.
  
For the sake of model comparison, the eccentricity of hydro evolution will be checked.  
The generalized space eccentricity of order $n$ reads \begin{equation}   
 \epsilon_{r,n} e ^ {i n \Phi_{r,n} } = \frac { <r^n e ^{i n \phi}>}{ < r^n>}, 
\end{equation}   
here
 < ... > stands for an energy density weighted space integral.

The momentum eccentricity provides a picture of the dynamic buildup of the elliptic flow of bulk hadrons and thermal photons.
The generalized momentum eccentricity of order $n$ reads \begin{equation}   
\epsilon_{p,n} e ^ {i n \Phi_{p,n} } = \frac { <v^n e ^{i n \phi_v }>}{ < v^n>}, \end{equation}   
again < ... > stands for an energy density weighted space integral.

The mean transverse flow $v_r$ is defined as energy density weighted space integral of the tranverse flow velocity
 \begin{equation} 
 v_r  = \frac { \int \epsilon  \sqrt{v_x^2 +v_y^2} d^3 x  } { \int  \epsilon   d^3 x}, 
\end{equation}   
where energy density $ \epsilon$  and  flow velocity $v_x,  v_y $ 
are provided by the hydrodynamic evolution of the collision system. 

Because of Lorentz transformations between the two frames, 
bigger $v_r$ can make the $p_{\rm t}$ spectrum of thermal photons harder, 
which means the temperature extracted according to the inverse slope of the $p_{\rm t}$ spectrum is higher than the temperature of the fluid where photons are emitted.
Thus, bigger $v_r$ can enlarge the fraction of thermal photons and increase the flow of direct photons according to eq.~(\ref{normalize}).

The transverse flow $v_r$ itself is also important to make big thermal photon flows $v_n^{\rm th}$.
We know that the momentum eccentricity provides a picture of the dynamic buildup of the elliptic flow of thermal photons.
The momentum eccentricity of order 2, $\epsilon_{p,2}$, is noted as $v^{\rm hydro}_2$ in our previous study~\cite{Liu:2009kta}.
In this study, Glauber initial condition makes a monotonous increase of $\epsilon_{p,2}$ for Au+Au collisions at 
$\sqrt {s_{\rm NN}}$ with centrality varying from 0-10\% to 60-70\%. 
But the $v_2$ of thermal photons does not increase monotonously. 
It decreases from 40-50\% to 60-70\% centrality because of the weaker and weaker $v_r$.   

\section{Results and analysis}
\begin{figure*}
\includegraphics[scale=0.8]{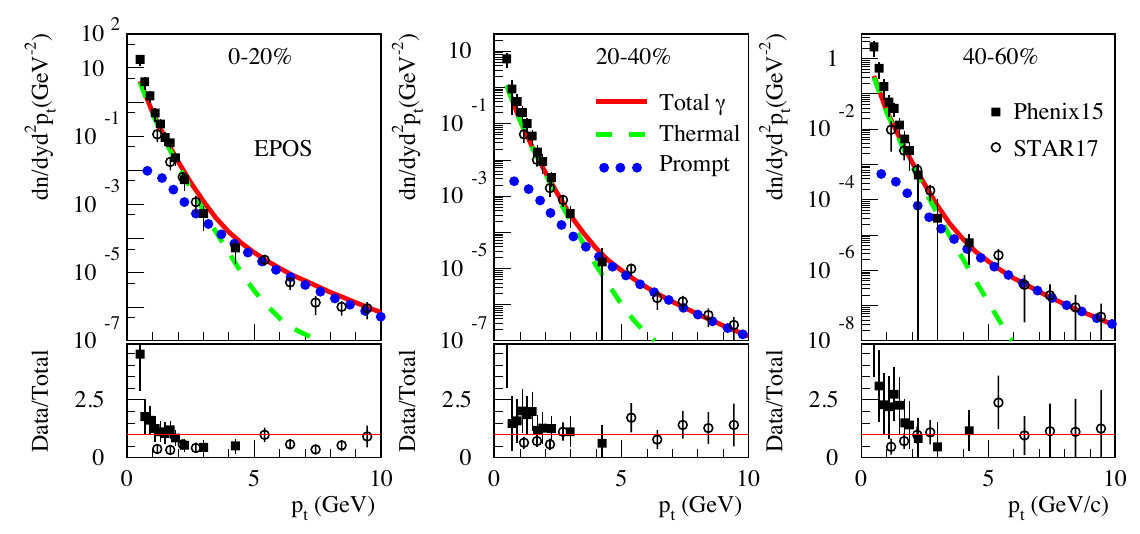}

 \caption{\label{200spectra} (Color Online) The transverse momentum spectra of 
prompt photons (dotted lines),  thermal photons (dashed lines) and direct photons (solid lines) from AuAu collisions 
at $\sqrt{s_{NN}}=200$~GeV for centrality 0-20\%, 20-40\% and 40-60\% 
calculated with EPOS initial condition are compared to
STAR~\cite{STAR:2016use} and PHENIX~\cite{PHENIX:2015igl}
 data. }   
\end{figure*}

We started with the $p_{\rm t}$ spectra calculated with EPOS initial condition, in Fig.~\ref{200spectra}.
The results of direct photons (solid lines), prompt photons (dotted lines), and thermal photons (dashed lines)
 from AuAu collisions at $\sqrt{s_{NN}}=200$~GeV for centrality  0-20\%, 20-40\% and 40-60~%
are compared with latest experimental data.  
Prompt photons dominate high $p_{\rm t}$ region and agree quite well with the STAR data~\cite{STAR:2016use} (empty dots)
 for all three centralities. 
Thermal photons dominate low $p_{\rm t}$ region and almost agree with the PHENIX data~\cite{PHENIX:2015igl} (full squares). 

The spectra at $p_{\rm t} < $0.5~GeV/c are evidently lower than data for all three centralities.
This $p_{\rm t}$ region is beyond the power of present calculation which is obtained  based on perturbative theory and the approximation that 
the energy of emitted photons is  far greater than the temperature of collision systems.

\begin{figure*}
\includegraphics[scale=0.8]{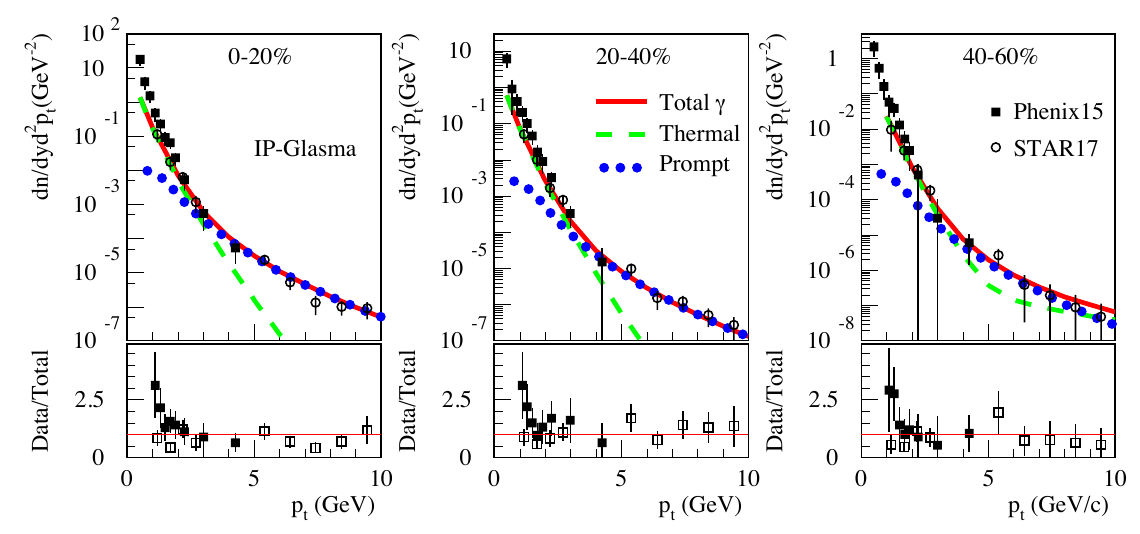}
 \caption{\label{200spectra-2} (Color Online) 
Same as Fig~\ref{200spectra} but calculated with IP-Glasma initial condition.  }
\end{figure*}

The $p_{\rm t}$ spectra calculated with IP-Glasma initial condition are shown in Fig.~\ref{200spectra-2}. 

We first compare them with published results shown in Fig.~6 of \cite{Gale:2021emg}.  
Pre-equilibrium emission is ignored by us, but considered in \cite{Gale:2021emg}, 
where this contribution never dominates, lower than thermal photons at low $p_{\rm t}$, and lower than prompt photons at high $p_{\rm t}$. 
So the comparison can be easily realized via the Data/Total ratio in the lower panels. 
The comparison focuses on thermal photons. 
At low pt region, where thermal photons dominate, our results are consistent with the published one, 
with ignorable difference due to the correction to emission rate, pre-equilibrium emission, chemical equilibrium time and so on.  
We notice that the $p_{\rm t}$ spectrum of thermal photons for centrality 40-60\% 
is evidently very hard, especially at the region of $p_{\rm t}>5~Gev/c$.
We note this as an upward leg in $p_{\rm t}$-spectra and discuss later.

Now we compare the $p_{\rm t}$ spectra calculated by us with two initial conditions. 
The $p_{\rm t}$ spectra from two initial conditions are quite similar in general, except the upward leg for centrality 40-60\%. 

\begin{figure*}
\includegraphics[scale=0.8]{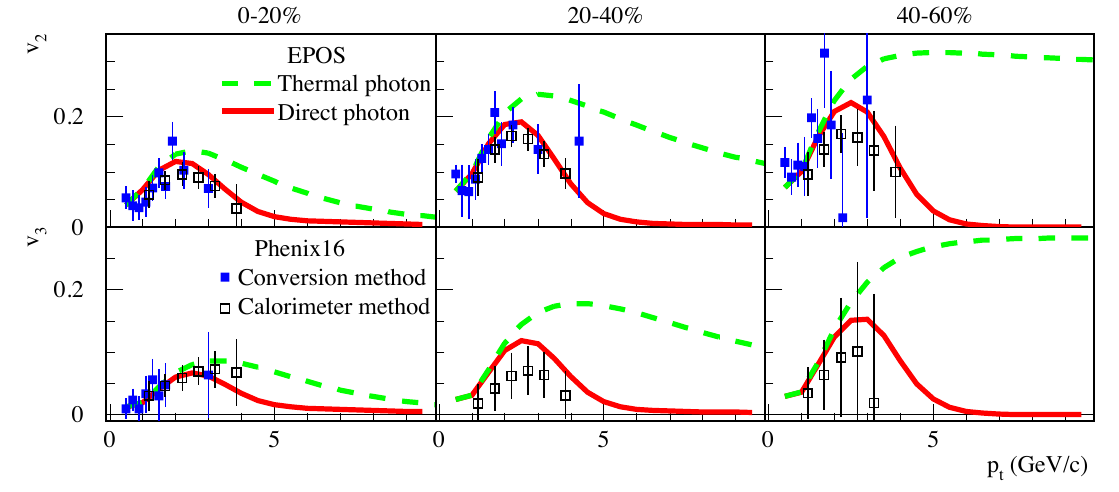}
 \caption{\label{v2v3} (Color Online) Elliptic flow $v_2$ (upper panels) and triangular flow $v_3$(lower panels) of direct photons (solid lines) and thermal photons (dashed lines) from AuAu collisions at $\sqrt{s_{NN}}=200$~GeV 
 for centrality 0-20\% , 20-40\% and 40-60\% calculated with EPOS initial condition. 
Data points $v_2$ and $v_3$ of direct photons from PHENIX\cite{PHENIX:2015igl} (squares) . }
\end{figure*}

In Fig.~\ref{v2v3} the calculated elliptic flow $v_2$ (upper panels) and triangular flow $v_3$(lower panels) of direct photons (solid lines) 
 and thermal photons (dashed lines) from AuAu collisions at $\sqrt{s_{NN}}=200$~GeV 
 for centrality 0-20\% , 20-40\% and 40-60\% with EPOS initial condition are shown. 
Latest data points $v_2$ and $v_3$ of direct photons from PHENIX\cite{PHENIX:2015igl} (squares) are shown as well. 
The elliptic flow $v_2$ of direct photons for all three centralities in this calculation coincide with experimental data! 
The $v_3$ of direct photons coincide with data for centrality 0-20\%, and slightly higher than data for centrality 20-40\%. For centrality 40-60\%, the calculated curve meets the large error bars of data.
The calculated $v_2$ and $v_3$ differ very little in $p_{\rm t}$ dependence and centrality dependence, 
as the system evolution is governed by the unitary hydrodynamics. 
The measured $v_2$ has an evident increase with centrality, but $v_3$ doesnot, 
if the error bars are shorten with better measurements. 
Then physics reason beyond hydrodynamics, ie, the EM fields, will show up, and   
a simultaneous explain to $v_2$ and $v_3$ within the unitary hydrodynamics will become an extravagant expectation.

Come back to the hydro calculation. Why do not the $v_n$ of thermal photons increase with $p_{\rm t}$ monotonously? Why $v_n$ show such a centrality dependence? 
The $p_{\rm t}$ spectra, $v_2$ and $v_3$ of thermal photons from AuAu collision of 40-60\% centrality in Fig.~\ref{200spectra} and Fig.~\ref{v2v3} are decomposed in Fig.~\ref{cent60ta_new} according to emission time. 
The time decomposition in Fig.~\ref{cent60ta_new} (EPOS initial condition) and Fig.~\ref{IP4060ta_new} (IP-Glamsma initial condition) is a direct presentation the hydro evolution of collision systems shown in Fig.~\ref{Fig-IP} via thermal photons.
We will explain the obtained $v_n$ of thermal photons and direct photons after introducing them.

 \begin{figure*}
\includegraphics[scale=0.8]{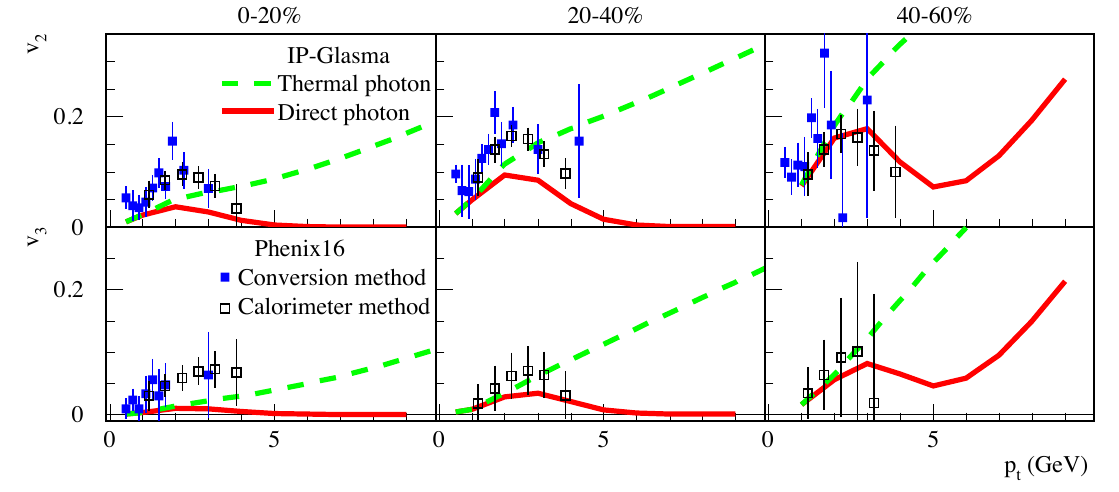}
\caption{\label{v2v3IP} (Color Online) 
Same as Fig~\ref{v2v3} but calculated with IP-Glasma initial condition.  }
\end{figure*}

In Fig.~\ref{v2v3IP} are shown elliptic flow $v_2$ (upper panels) and triangular flow $v_3$(lower panels) calculated with IP-Glasma initial condition. 
The calculated elliptic flow $v_2$, and also $v_3$ are lower than PHENIX data for centrality 0-20\% and 20-40\%,
consistent with published result\cite{Gale:2021emg}.
However, the calculated $v_2$ and $v_3$ coincide with data for centrality 40-60\%, different from \cite{Gale:2021emg}.

We come back to the upward leg mentioned in Fig.~\ref{200spectra-2} for this centrality.
This is due to the existence of hot spots in the collision system. 
Upward legs of $p_{\rm t}$ spectra do not appear if the energy densities of hot spots decrease gradually. 
However, hot spots may have a cliff-style decline of energy density due to viscous hydrodynamics.
The effect of viscosity in smaller systems is stronger, because a bigger pressure gradient produces a strong deviation from mechanical equilibrium, ie, hot spots as subsystems.
The cliff-style decline of hot spots makes upward legs appear in $p_{\rm t}$ spectra, with more details shown in the time decomposition in Fig.~\ref{IP4060ta_new}, at $\tau =$0.4, 0.9, and 1.4~fm/c. 

Viscous correction of photon emission rate may smooth $p_{\rm t}$ spectra and make upward legs disappear.
The absence of viscous correction to photon emission rate makes our IP-Glasma calculation differ from   \cite{Gale:2021emg} 
in peripheral collisions, stronger at high $p_{\rm t}$ region.
The upward leg of $p_{\rm t}$ spectrum increases the fraction of thermal photons in eq.~\ref{normalize}, thus bigger $v_n$ than \cite{Gale:2021emg} for more peripheral collisions. 
It provokes the competition between thermal and prompt photon at high  $p_{\rm t}$ region. 
So a second increase of $v_n$ of direct photons in AuAu collisions with 40-60\% centrality show up in Fig.~\ref{v2v3IP}.

The second increase of  $v_2$ and $v_3$ direct photon  in AuAu collisions with 40-60\% centrality is at quite high $p_{\rm t}$.
$v_n$ at this $p_{\rm t}$ region is not measured and not plotted in previous work~\cite{Gale:2021emg}. 
This measurement may offer a good constraint of the viscous correction to photon emission rate, which by now nobody has done  perfectly.

Hot spots with cliff-style decline also appear in EPOS at initial time $\tau_0$ in AuAu collisions with centrality 0-20\%, 
the most central collisions, because of the superposition of a big number pomerons. 
A upward leg in the $p_{\rm t}$ spectrum shows up (blinks)  at the moment.

 \begin{figure*}
\includegraphics[scale=0.8]{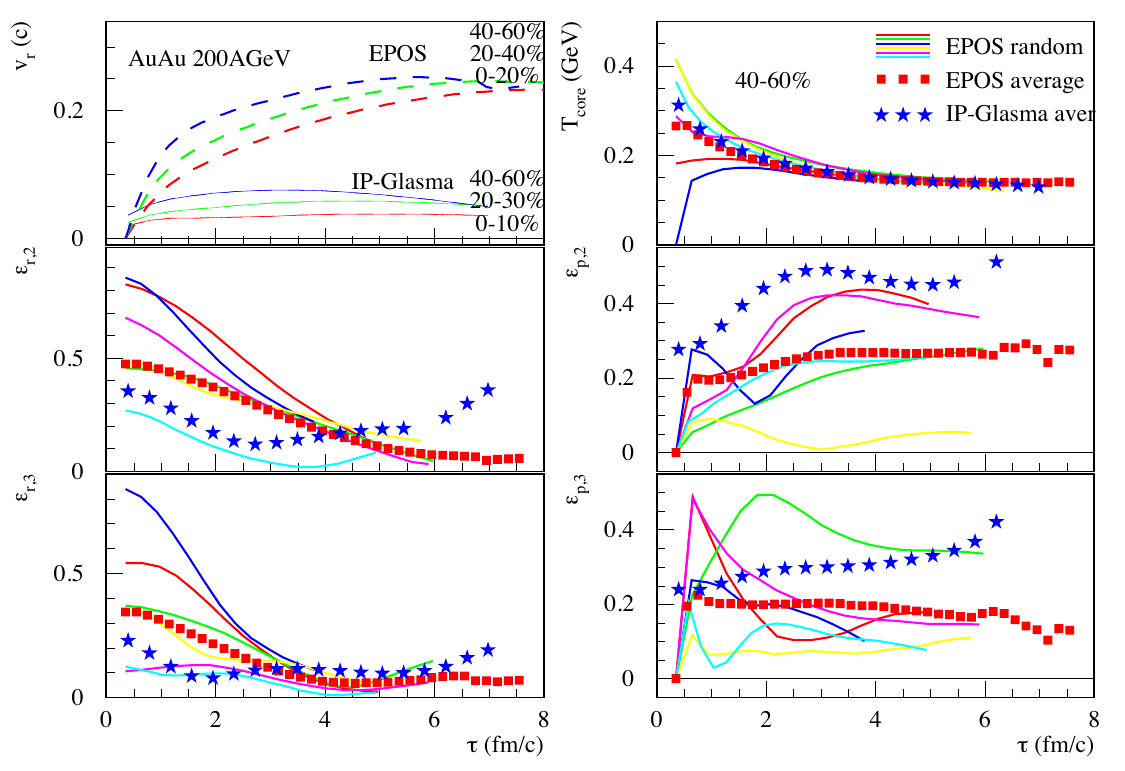}
\caption{\label{Fig-IP} (Color Online) 
Comparison the evolution from EPOS and IP-Glasma initial conditions.  
Successively the mean transverse flow $v_r$, 
the temperature at the center point with Cartesian coordinates (x,y,z)=(0,0,0), 
space eccentricity and momentum eccentricity of order 2 and 3 are plotted as a function of proper time $\tau$.
Centrality dependence is shown in first panel. In the other panels,
collision centrality focus on 40-60\%, with the space to show a few random EPOS events.  }
\end{figure*}

In Fig.~\ref{Fig-IP} successively are shown the time evolution of the mean transverse flow $v_r$, 
the temperature at the center point, space eccentricity and momentum eccentricity of order 2 and 3, 
for AuAu collisions with centrality 40-60\% from both initial conditions.

In the first panel, the centrality dependence of $v_r$ is also shown. 
The same centrality dependence is obtained from the two initial conditions, EPOS (dashed lines) and IP-Glasma (solid lines):  
$v_r$ increase, from central to peripheral collisions. 
The reason is that a smaller size of the collision system produces bigger pressure gradients to push the collective motion and gain bigger $v_r$.
We notice that $v_r$ has a different centrality dependence 
with the employment of ideal hydrodynamics \cite{Liu:2009kta}. 

The initial flow velocity is zero in EPOS, so all dashed lines start from the same point ($\tau=0.35~fm/c, v_r=0$).  
But soon, $v_r$ in EPOS case are much bigger than IP-Glasma. 

In the other panels, we focus on centrality 40-60\%, where full squares stand for average results from EPOS initial conditions,
and stars for average results from IP-Glasma initial conditions. 
A few random EPOS events are also shown, and each event keeps the same color in the five panels.

In the second panel is the evolution of temperature at the center point with Cartesian coordinates (x,y,z)=(0,0,0). 
Generally the temperature decreases with time in both models. 
Energy density in both models agree with each other, on average.

In EPOS, very often, the center point is not the hottest for 40-60\% centrality, due to Pomeron distribution.  
This point is heated by the hot dense matter nearby, seeing the random event of the blue curve in this panel. 
There are more details on initial conditions, especially for EPOS case in \cite{Werner:2010aa}.

Eccentricity of 2nd order and 3rd order are shown in line 2 and line 3 of Fig.~\ref{Fig-IP}. 
The space eccentricity (left), is comparable on average, 
because both EPOS and IP-Glasma are related to Glauber distribution.   
But momentum eccentricity (right) differs a lot in the two models, 
due to different dynamical details in initial conditions.  
For example, the $\epsilon_{p,2}$ and $\epsilon_{p,3}$, from IP-Glasma (stars) are bigger than from EPOS (squares).

When hadrons emitted from the freeze-out supersurface, Lorentz transformations with flow velocity are also needed.
Thus the momentum eccentricity makes anisotropic hadron production.
How can both models explain successfully the anisotropic production of hadrons, ie, $v_2$ of pions, 
with so different momentum eccentricity? 
The reason is that the large transverse flow vecolity $v_r$ in EPOS (shown in the first panel of this figure) compensates
its weaker momentum eccentricity.
Therefore transverse flow plays an important role, not only for photons but also for hadrons.
A balance between eccentricity and flow velocity magnitude is required by both hadron and photon data. 

For a better understanding the results of thermal photons in Fig.~\ref{200spectra}-\ref{v2v3IP}, 
we decompose the contribution of thermal photons emitted from AuAu collisions with centrality 40-60\% according to the emission time.
The strong emission occurs at early time, thus the decomposed results in first 6~fm/c from $\tau_0$ are shown in Fig.~\ref{cent60ta_new}
 (EPOS initial condition) and Fig.~\ref{IP4060ta_new} (IP-Glasma initial condition) from left to right, 
where time interval is 0.5~fm/c, with color order: red, green, blue and yellow for the three columns, and $p_{\rm t}$ spectra (first line), 
$v_2$ (second line) and $v_3$ (third line). Now it is easy to check the response to the hydro evolution of collision system shown in Fig.~\ref{Fig-IP}. 

For EPOS, the $p_{\rm t}$-spectrum at $\tau$ =0.35~fm/c is hard due to high energy density from pomeron superposition.
In the later time, $p_{\rm t}$ spectra change little with time, because the increase of flow velocity $v_r$ compensates the
decrease of temperature.
For IP-Glasma, $v_r$ is almost constant, the $p_{\rm t}$ spectra show the decrease of temperature neatly, 
except hot spots make three upward legs as explained above.

$v_2$ and $v_3$ are effected by both momentum eccentricity and flow velocity magnitude.  
For example, in the first 2~fm/c, there is an increase of $v_r$ in EPOS, and increase of momentum eccentricity in IP-Glasma, 
so $v_2$ and $v_3$ increase with time evidently in both models.
In the later time, $v_r$ and momentum eccentricity vary little, so do $v_2$ and $v_3$ of thermal photons, in both models. 

Here we see, $v_2$ and $v_3$ of thermal photons emitted at any $\tau$ increase with $p_{\rm t}$ monotonously. 
Thus, in Fig.~\ref{v2v3IP},  $v_2$ and $v_3$ of thermal photons from IP-Glasma increase  with $p_{\rm t}$ monotonously as well.
But why so different is the $p_{\rm t}$ dependence of thermal photon $v_n$ in EPOS, shown as Fig.~\ref{v2v3} ? 
What makes the centrality dependence ---- $v_n$ of thermal photons decrease more rapid in more central collisions? 
The reasons are following.

First, the initial flow velocity is set as zero in EPOS, which makes $v_n=0,n=2,3$ for thermal photons emitted at this time interval,
shown as red lines which coincide with the horizontal axis 
in  the first column of  Fig.~\ref{cent60ta_new}.

Second, the $p_{\rm t}$-spectrum at $\tau$ =0.35~fm/c is hard, much hard than later time, as shown in first column   
of   Fig.~\ref{cent60ta_new}, due to high energy density from pomeron superposition. 
The emission of the first time interval makes a big fraction in thermal contribution.
Comparing with emission of later time,  the fraction increases evidently with $p_{\rm t}$. 
The more this fraction dominates thermal emission, the lower $v_n$ of thermal photons obtained. 
The two factors make the $p_{\rm t}$ dependence of $v_n$ of thermal photons.

The pomeron number increases with centrality, and their superposition makes hot spots at $\tau_0$ =0.35~fm/c more possible.
This also increases the fraction of emission in the first time interval, and suppresses the  $v_n$ of thermal photons.
This gives the  centrality dependence of $v_n$ of thermal photons in EPOS.
With a further suppression from prompt photons according to eq.~\ref{normalize}, we get the $v_n$ of direct photons as shown in Fig.~\ref{v2v3}.

\begin{figure*}
\includegraphics[scale=0.8]{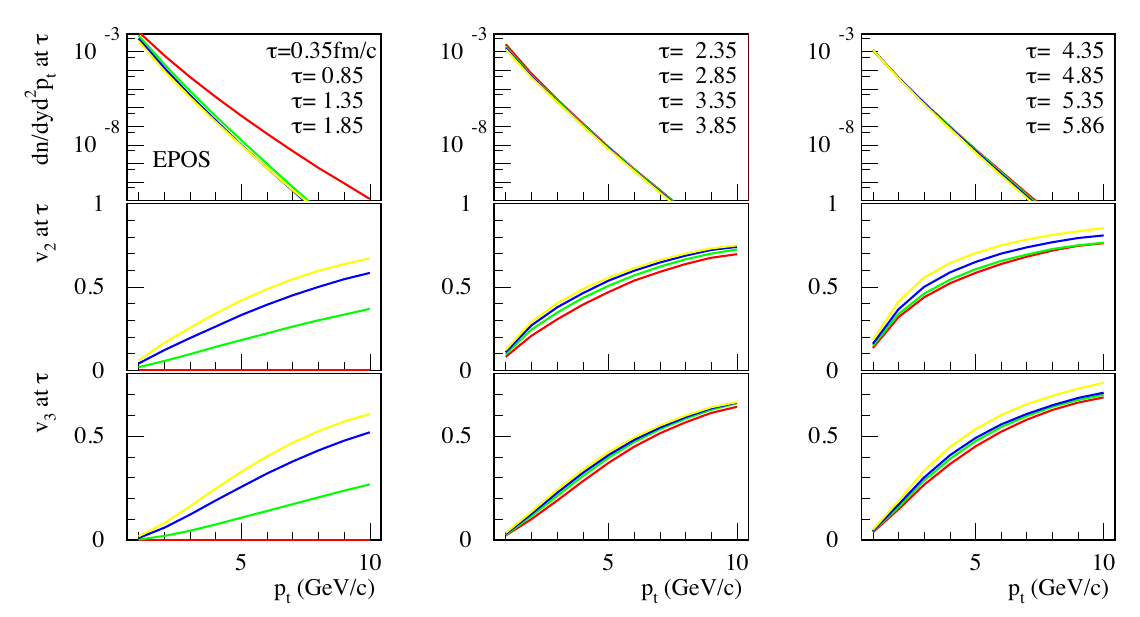}
 \caption{\label{cent60ta_new} (Color Online)
The time decomposition of $p_{\rm t}$ spectrum (first line), $v_2$ (second line) and $v_3$ (third line) 
of thermal photons emitted from AuAu collisions at $\sqrt {s_{\rm NN}}$=200~GeV at centrality 40-60\%.
Time interval is 0.5~fm/c, with color order: red, green, blue and yellow for the three columns.  
Calculated with EPOS initial condition. 
}
\end{figure*}

\begin{figure*}
\includegraphics[scale=0.8]{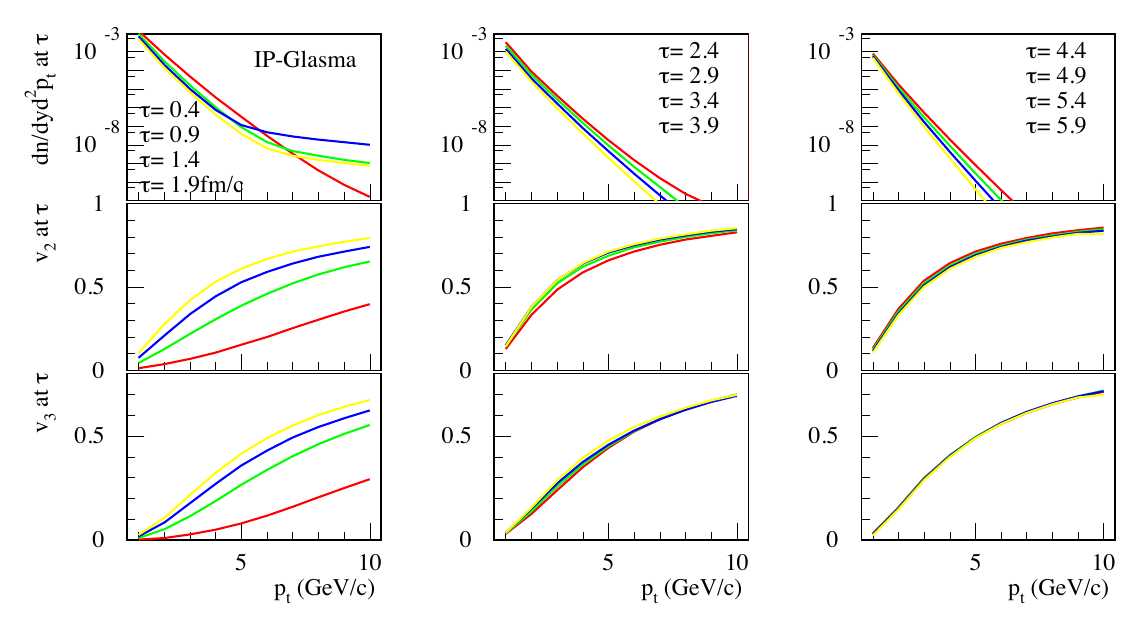}
 \caption{\label{IP4060ta_new} (Color Online) 
Same as Fig.~\ref{cent60ta_new}, but calculated with IP-Glasma initial condition. } 
\end{figure*}

\begin{figure*}
\includegraphics[scale=0.8]{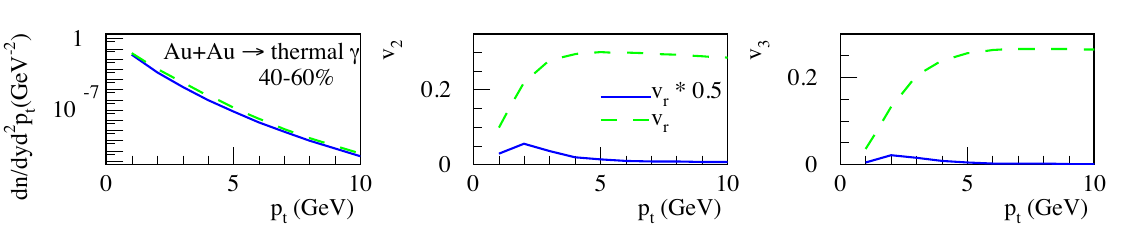}

 \caption{\label{b60} (Color Online)   
The dashed lines are the EPOS normal results of AuAu collisions at $\sqrt{s_{NN}}=$200 GeV with 40-60\%. 
Solid lines are the results with reduced 50\% flow velocity. }
\end{figure*}

To show how strong is the effect from flow velocity, we recalculated photon emission from AuAu collisions 
at centrality 40-60\% with EPOS initial condition, where flow velocity is reduced to 50\%, but energy density and eccentricity kept.
In Fig~\ref{b60}, the $p_{\rm t}$-spectrum, $v_2$ and $v_3$ of thermal photons from AuAu collision at 40-60\%  shown again, as dashed lines.
The results from the reduced flow velocity are shown as solid lines.
The reduction of flow velocity does not make an evident modification in  $p_{\rm t}$ spectrum,
but it can suppress strongly $v_2$ and $v_3$ of thermal photons.

\section{ CONCLUSIONS and DISCUSSION}
We investigated direct photon production in AuAu collisions at $\sqrt {s_{\rm NN}}$=200~GeV with two photon sources, prompt photons and thermal photons. Thermal photons were calculated with 3+1D viscous hydrodynamical description of the collision systems, with two types of initial conditions, EPOS and IP-Glasma.

Take into account of the uncertainty of the system evolution, we ignored the viscous correction to the emission rate, which seems to work well in general. The calculated $p_{\rm t}$ spectra of direct photons from both initial conditions agree quite well with latest direct photon data, for all centralities.

A upward leg of $p_{\rm t}$ spectrum appears due to the cliff-style decline of hot spots in viscous hydrodynamics and the absence of viscous correction to photon emission rate. The region is clear, $p_{\rm t} > $5~GeV/c, precisely, centrality 40-60\% AuAu collisions with IP-Glasma initial condition, and centrality 0-20\% AuAu collisions with EPOS initial condition.

Our IP-Glasma results coincide with published work in most cases, though the viscous correction to photon emission rate is treated differently.  The difference becomes visible in peripheral collisions (40-60\%) and bigger at higher $p_{\rm t}$, due to the upward leg of $p_{\rm t}$ spectrum.

The observed large $v_2$ of direct photons are nicely reproduced with EPOS initial condition for all centralities. The calculated $v_3$ of direct photon are also within the errors bar of experimental data.
We understand the results via comparing the hydro evolution from the two initial conditions and the time-decomposition of the obtained $p_{\rm t}$ spectra, $v_2$ and $v_3$.
We find that both the eccentricity and magnitude of flow velocity are important for the anisotropic direct photon production. With the same eccentricity, magnitude of flow velocity can change the emitted photon a lot. We remind the constraints to hydro models, ie, the balance between eccentricity and flow velocity, via not only hadron data, but via photon data as well.

We understand the puzzle of direct photons to some degree through this work, but there are still a lot unknown to us. 
In EPOS, it is an ansatz that initial flow velocity is zero. This ansatz is confirmed with direct photon data to some degree. 
Suppose a strong emission at early time from the collision system at the high temperature, and a non-zero flow velocity 
(with some eccentricity), it will produce a very big elliptic flow of direct photons.
At the other hand, the realization of thermal and chemical equilibrium in our intuition is gradually. The pre-equilibrium emission might be weak and ignorable. 
But zero flow velocity seems a pain for thermalization.

\begin{acknowledgments}
This work was supported by the Natural Science Foundation of China
under Project No.11275081 and by the Program for New Century Excellent
Talents in University (NCET).
 \end{acknowledgments}


\begin{thebibliography}{10}


\bibitem{PHENIX:2011oxq}
A.~Adare \textit{et al.} [PHENIX],
``Observation of direct-photon collective flow in $\sqrt{s_{NN}}=200$ GeV Au+Au collisions,''
Phys. Rev. Lett. \textbf{109}, 122302 (2012)
doi:10.1103/PhysRevLett.109.122302
[arXiv:1105.4126 [nucl-ex]].

\bibitem{PHENIX:2015igl}
A.~Adare \textit{et al.} [PHENIX],
Phys. Rev. C \textbf{94}, no.6, 064901 (2016)
doi:10.1103/PhysRevC.94.064901
[arXiv:1509.07758 [nucl-ex]].

\bibitem{David161}
 "Electromagnetic radiation from hot and dense hadronic matter," https://ectstar.fbk.eu/node/4229 (2018).

\bibitem{David309}
 Thermal photons and dileptons, https://www.bnl.gov/tpd/ (2011).

\bibitem{David310}
 [310] Thermal radiation workshop, https://www.bnl.gov/trw2012/ (2012).

 \bibitem{David311}
Electromagnetic probes of strongly interacting matter: Status and future of low-mass lepton-pair spec
troscopy, http://www.ectstar.eu/node/92 (2013).

 \bibitem{David312}
Emmi rrtf on direct-photon ow puzzle, https://indico.gsi.de/conferenceDisplay.py?confId=2661
 (2014).

\bibitem{David313}
 Thermal photons and dileptons in heavy-ion collisions, https://www.bnl.gov/tpd2014/ (2014).

\bibitem{David314}
 New perspectives on photons and dileptons in ultrarelativistic heavy-ion collisions at rhic and lhc, 
http://www.ectstar.eu/node/1232 (2015).

\bibitem{David:2019wpt}
G.~David,
``Direct real photons in relativistic heavy ion collisions,''
Rept. Prog. Phys. \textbf{83}, no.4, 046301 (2020)
doi:10.1088/1361-6633/ab6f57
[arXiv:1907.08893 [nucl-ex]].

\bibitem{Gale:2014dfa}
C.~Gale, Y.~Hidaka, S.~Jeon, S.~Lin, J.~F.~Paquet, R.~D.~Pisarski, D.~Satow, V.~V.~Skokov and G.~Vujanovic,
``Production and Elliptic Flow of Dileptons and Photons in a Matrix Model of the Quark-Gluon Plasma,''
Phys. Rev. Lett. \textbf{114}, 072301 (2015)
doi:10.1103/PhysRevLett.114.072301
[arXiv:1409.4778 [hep-ph]].
 
\bibitem{vanHees:2011vb}
H.~van Hees, C.~Gale and R.~Rapp,
``Thermal Photons and Collective Flow at the Relativistic Heavy-Ion Collider,''
Phys. Rev. C \textbf{84}, 054906 (2011)
doi:10.1103/PhysRevC.84.054906
[arXiv:1108.2131 [hep-ph]].

\bibitem{Linnyk:2013wma}
O.~Linnyk, W.~Cassing and E.~L.~Bratkovskaya,
``Centrality dependence of the direct photon yield and elliptic flow in heavy-ion collisions at $\sqrt{s_{NN}}=200$ GeV,''
Phys. Rev. C \textbf{89}, no.3, 034908 (2014)
doi:10.1103/PhysRevC.89.034908
[arXiv:1311.0279 [nucl-th]].

\bibitem{Chiu:2012ij}
M.~Chiu, T.~K.~Hemmick, V.~Khachatryan, A.~Leonidov, J.~Liao and L.~McLerran,
``Production of Photons and Dileptons in the Glasma,''
Nucl. Phys. A \textbf{900}, 16-37 (2013)
doi:10.1016/j.nuclphysa.2013.01.014
[arXiv:1202.3679 [nucl-th]].

\bibitem{Monnai:2014kqa}
A.~Monnai,
``Thermal photon $v_2$ with slow quark chemical equilibration,''
Phys. Rev. C \textbf{90}, no.2, 021901 (2014)
doi:10.1103/PhysRevC.90.021901
[arXiv:1403.4225 [nucl-th]].

\bibitem{Gale:2021emg}
C.~Gale, J.~F.~Paquet, B.~Schenke and C.~Shen,
``Multimessenger heavy-ion collision physics,''
Phys. Rev. C \textbf{105}, no.1, 014909 (2022)
doi:10.1103/PhysRevC.105.014909
[arXiv:2106.11216 [nucl-th]].

\bibitem{Jia:2022awu}
M.~Jia, H.~Li and D.~Hou,
``The photon production and collective flows from magnetic induced gluon fusion and splitting in early stage of high energy nuclear collision,''
Phys. Lett. B \textbf{846}, 138239 (2023)
doi:10.1016/j.physletb.2023.138239
[arXiv:2211.16770 [hep-ph]].

\bibitem{Ayala:2022zhu}
A.~Ayala, J.~D.~Casta\~no-Yepes, L.~A.~Hern\'andez, A.~J.~Mizher, M.~E.~Tejeda-Yeomans and R.~Zamora,
``Anisotropic photon emission from gluon fusion and splitting in a strong magnetic background: The two-gluon one-photon vertex,''
Phys. Rev. C \textbf{106}, no.6, 064905 (2022)
doi:10.1103/PhysRevC.106.064905
[arXiv:2209.09364 [hep-ph]].

\bibitem{Sun:2023rhh}
J.~A.~Sun and L.~Yan,
Phys. Rev. C \textbf{109}, no.3, 034917 (2024)
doi:10.1103/PhysRevC.109.034917
[arXiv:2311.03929 [nucl-th]].

\bibitem{Fujii:2022hxa}
H.~Fujii, K.~Itakura, K.~Miyachi and C.~Nonaka,
Phys. Rev. C \textbf{106}, no.3, 034906 (2022)
doi:10.1103/PhysRevC.106.034906
[arXiv:2204.03116 [nucl-th]].

\bibitem{Werner:2013tya}
K.~Werner, B.~Guiot, I.~Karpenko and T.~Pierog,
``Analysing radial flow features in p-Pb and p-p collisions at several TeV by studying identified particle production in EPOS3,''
Phys. Rev. C \textbf{89}, no.6, 064903 (2014)
doi:10.1103/PhysRevC.89.064903
[arXiv:1312.1233 [nucl-th]].

\bibitem{Werner:2010aa}
K.~Werner, I.~Karpenko, T.~Pierog, M.~Bleicher and K.~Mikhailov,
``Event-by-Event Simulation of the Three-Dimensional Hydrodynamic Evolution from Flux Tube Initial Conditions in Ultrarelativistic Heavy Ion Collisions,''
Phys. Rev. C \textbf{82}, 044904 (2010)
doi:10.1103/PhysRevC.82.044904
[arXiv:1004.0805 [nucl-th]].

\bibitem{Schenke:2012hg}
B.~Schenke, P.~Tribedy and R.~Venugopalan,
``Event-by-event gluon multiplicity, energy density, and eccentricities in ultrarelativistic heavy-ion collisions,''
Phys. Rev. C \textbf{86}, 034908 (2012)
doi:10.1103/PhysRevC.86.034908
[arXiv:1206.6805 [hep-ph]].


\bibitem{Schenke:2012wb}
B.~Schenke, P.~Tribedy and R.~Venugopalan,
``Fluctuating Glasma initial conditions and flow in heavy ion collisions,''
Phys. Rev. Lett. \textbf{108}, 252301 (2012)
doi:10.1103/PhysRevLett.108.252301
[arXiv:1202.6646 [nucl-th]].


\bibitem{Paquet:2015lta}
J.~F.~Paquet, C.~Shen, G.~S.~Denicol, M.~Luzum, B.~Schenke, S.~Jeon and C.~Gale,
Phys. Rev. C \textbf{93}, no.4, 044906 (2016)
doi:10.1103/PhysRevC.93.044906
[arXiv:1509.06738 [hep-ph]].

\bibitem{Gale:2012rq}
C.~Gale, S.~Jeon, B.~Schenke, P.~Tribedy and R.~Venugopalan,
Phys. Rev. Lett. \textbf{110}, no.1, 012302 (2013)
doi:10.1103/PhysRevLett.110.012302
[arXiv:1209.6330 [nucl-th]].


\bibitem{Owens1987}J.F.~Owens, Rev. Mod. Phys. 59, 465 (1987).

\bibitem{Liu:2008eh}
F.~M.~Liu, T.~Hirano, K.~Werner and Y.~Zhu,
``Centrality-dependent direct photon p(t) spectra in Au + Au collisions at RHIC,''
Phys. Rev. C \textbf{79}, 014905 (2009)
doi:10.1103/PhysRevC.79.014905
[arXiv:0807.4771 [hep-ph]].

\bibitem{Martin:2001es} 
A.~D.~Martin, R.~G.~Roberts, W.~J.~Stirling and R.~S.~Thorne, 
``MRST2001: Partons and $\alpha_s$ from precise deep inelastic scattering and Tevatron jet data,''
 Eur.\ Phys.\ J.\ C \textbf{23}, 73 (2002) [arXiv:hep-ph/0110215].

\bibitem{AMY}P.~Arnold, G.~D.~Moore, and L.~G.~Yaffe, J.~High
Energy Phys. \textbf{0111}, 057 (2001); J. High Energy Phys. \textbf{0112},
9 (2001).

\bibitem{MYM} S.\ Turbide, R.\ Rapp and C.\ Gale, Phys.\ Rev.\ C
\textbf{69}, 014903 (2004).

\bibitem{NEXUS} H.J. Drescher, M. Hladik, S. Ostapchenko, T. Pierog, K. Werner,
" Parton based Gribov-Regge theory",
    hep-ph/0007198, Phys.Rept. 350 (2001) 93-289.

\bibitem{VENUS} K. Werner,
" Strings, pomerons. and the VENUS model of hadronic interactions at ultra-relativistic energies",
    Phys. Rept. \textbf{232}, 87 (1993).

\bibitem{Liu:2009kta} 
  F.~M.~Liu, T.~Hirano, K.~Werner and Y.~Zhu,
  ``Elliptic flow of thermal photons in Au + Au collisions at s(NN)**(1/2) = 200-GeV,''
  Phys.\ Rev.\ C {\bf 80}, 034905 (2009)
  doi:10.1103/PhysRevC.80.034905
  [arXiv:0902.1303 [hep-ph]].

\bibitem{STAR:2016use}
L.~Adamczyk \textit{et al.} [STAR],
Phys. Lett. B \textbf{770}, 451-458 (2017)
doi:10.1016/j.physletb.2017.04.050
[arXiv:1607.01447 [nucl-ex]].
 

\end{thebibliography}
\end{document}